\documentstyle[aps]{revtex}
\begin{document}
\bibliographystyle{prsty} 

\wideabs{
\title{ Potential of an ionic impurity in a large $^4$He cluster}
\author{ K.~K.~Lehmann\footnote{1998 Visiting Fellow.  
Permanent address: Department of chemistry,
Princeton University, Princeton NJ 08544}}
\address{JILA, University of Colorado and National Institute of Standards
and Technology, Boulder, CO 80309--0440}
\author{Jan A. Northby}
\address{Physics Department, University of Rhode Island, Kingston, RI 02881}

\date {To be published in \em Molecular Physics}

\maketitle 

\begin{abstract} 
This paper presents an analysis of the motion of an impurity ion
in a nanometer scale $^4$He cluster.  Due to induction forces,
ions are strongly 
localized near the center of
the cluster, with a root mean squared thermal displacements of only a 
few \AA.  The trapping potential is found to be nearly harmonic,
with a frequency of $2.3(1.0)\,$GHz for a positive (negative) ion
in a He cluster of radius $5\,$nm.  
The anharmonicity is small and positive (energy increases slightly
faster than linear with quantum number).  It is suggested that by using
frequency sweep microwave radiation, it should be possible to drive
the ion center of mass motion up to high quantum numbers, allowing
the study of the critical velocity as a function of cluster size.

\end{abstract}

}
\section{Introduction}

The last few years have seen dramatic advances in the spectroscopy of
atoms and molecules attached to large He clusters~\cite{Whaley98}. 
These clusters provide a unique environment for a spectroscopy
which combines many of the attractive features of both
high resolution gas phase spectroscopy and traditional
matrix spectroscopy~\cite{Lehmann98}.  
These include the ability to obtain rotationally resolved spectra of
even very large molecules~\cite{Hartmann_thesis}, 
and the ability to form and stabilize
extremely fragile species~\cite{Higgins96a}, 
including high spin states of molecules.  

Despite rapid progress, many fundamental questions remain about
spectroscopy in this environment.
One important topic that has received little attention is the
dynamics of the center of mass motion of an impurity in a
$^4$He cluster~\cite{Toennies95}.  
A recent experiment reported by Grebenev, 
Toennies, and Vilesov~\cite{Grebenev98} has convincingly demonstrated
that these clusters are superfluid.  Thus we expect that an 
impurity, like in bulk superfluid He, should be able to move
with little or no friction as long as its velocity stays below
a `critical velocity' which is found to be on the order of 
$30\,$m/s for motion of positive ions in bulk superfluid 
$^4$He~\cite{Rayfield64,McClintock95}.  
Doped He clusters provide an attractive system to study
the size dependence of superfluid hydrodynamics.

There have been several theoretical studies reported that consider the
motion of an electron bound to the surface of He or other dielectric
clusters~\cite{Nabutovskii85,Antoniewicz82,Ballester86,Krishna88,Rosenbilt94a,%
Rosenbilt94b}.
Most of these have only considered the electrostatic potential
for the ion outside of the cluster, and solved for the quantum levels of
the electron, including determining the minimum size for a cluster to
bound an electron.  
The only exceptions in the open literature known to the authors is the work of 
Antoniewicz {\it et al.}~\cite{Antoniewicz82,Ballester86} that presents the
electrostatic potential inside the cluster, but the potential given in that
work is in error by a factor of two, and a paper by Northby, Kim 
and Jiang~\cite{Northby94b} that presents an approximate potential, similar
to that given in section 2 below, but without any derivation.  
The exact electrostatic expressions have been given in the Ph.D. thesis
of Ballester~\cite{Ballester_thesis} and Kim~\cite{Kim_thesis}.

This paper will present a realistic potential for the motion
of an ion in a He cluster, based upon the electrostatic
potential produced by the dielectric response of the He to
the ion.  Similar response is partly responsible for the 
trapping potential of an ion beneath a He--vapor interface
that was exploited by Poitrenaud and Williams~\cite{Poitrenaud72}
to determine the effective mass of positive and negative
charge carriers in bulk He.  This paper will then propose
experiments that can be viewed as the natural extension of
this earlier work on bulk He.  In addition to allowing the 
determination of the size dependence of the effective mass
of ions, it should also allow for the study of the effective size
dependence of critical velocity of ion in the cluster.  

In a following paper, the motion of a neutral impurity 
atom or molecule will be considered, and an effective
Hamiltonian for its motion, including both
long range potential and hydrodynamic contributions,
will be derived.  The potential localizes the neutral 
impurity near the center of the cluster, though much more
weakly than for the case of an ion which has much stronger
long range interactions with the He, falling as
$r^{-4}$ versus $r^{-6}$.  
In the case of molecules, there is
a coupling of the rotation and the center of mass motion
which leads to a broadening mechanism in the 
rotational or ro--vibrational spectrum.  

\section{Potential for an Ion in a He droplet}

A large fraction of the studies of impurities in He clusters have
exploited mass spectrometry~\cite{Gspann78,Lewerenz93,Lewerenz95}.  
The migration of the charge in
a cluster (which is most likely initially localized on a
He atom) will be influenced by the effective potential of such a 
charge~\cite{Callicoatt96}.
By electrostatics, the charge will be most stable at the center of the
cluster, furthest from the polarization charge that will develop on the
cluster surface, due to the dielectric constant of He.
The purpose of this section is to derive an expression for this
potential and explore some of its predicted consequences.
This potential does not include the energy of solvation of the
He around the impurity.  

Assume that we have an ion of unit charge, $e$, 
at a radius $a$ from the center of droplet, and  pick the coordinate
system such that the $z$ axis is along the displacement of the ion from the 
center.  In order to calculate the energy, we can sum up the ion induced
dipole interaction of the ion with the `missing' He that would be `outside'
the droplet.  This makes the `zero' of energy an ion in an infinite bulk
of liquid He, and also avoids the difficulties with the nearby He atoms
which are strongly bound to the ion.

Let $r(\theta)$  be the distance from the ion to the droplet surface
at a polar angle $\theta$ measured {\it from the ion}.  
Basic trigonometry gives:
\begin{equation}
R^2 = (a + r \cos(\theta) )^2 + r^2 \sin(\theta)^2 = 
a^2 + r^2 + 2 ra \cos(\theta)
\end{equation}
from which we can derive:
\begin{equation}
r(\theta) = \sqrt{R^2 - a^2 \sin(\theta)^2} 
- a \cos(\theta) \label{eq:r(theta)}
\end{equation}
The field from a charge is given by:
\begin{equation}
E(r) = \frac{e}{4\, \pi\, \epsilon\, r^2}
\end{equation}
This leads to an energy difference from the bulk:
\begin{equation}
\Delta E = \int_0^{\pi}  \int_{r(\theta)}^{\infty}
\frac{1}{2}(\rho\, \alpha) E^2(r') \, 2\, \pi\, r'^2\, \sin(\theta) 
dr'\, d\theta 
\end{equation}
where $\rho$ is the number density of He 
($0.0218\, {\rm \AA}^{-3}$~\cite{Brink90})
 and $\alpha$ is the polarizability of He.
Evaluating the integrals (which were done using the
Mathcad program~\cite{Mathcad}), gives the following result:
\begin{equation}
\Delta E = \left(\frac{e^2}{4\,\pi\,\epsilon\, R}\right) 
\left(\frac{\rho\,\alpha}{4\,\pi\,\epsilon}\right) (2\,\pi) F_0(a/R)
\label{eq:delE_ion}
\end{equation}
with:
\begin{equation}
F_0(y) = \frac{1}{4} \left( \frac{2}{1-y^2} + \frac{1}{y} \log \left(
\frac{1+y}{1-y} \right) \right)  \label{eq:F_0}
\end{equation}
\begin{equation}
F_0(y) \approx 1 + \frac23y^2 + \frac35y^4 + \ldots
\end{equation}
In this paper, we will consistently use $y$ for the
reduced or fractional radius of the impurity ion or molecule.
Using the relationships between the polarizablity,
the electric susceptibility $\chi$, and the relative dielectric constant 
$\epsilon_r$: $\rho \alpha = \epsilon_0 \chi$ for $\chi \ll 1$,
$\chi = \epsilon_r -1$, and 
$\epsilon = \epsilon_r \epsilon_0$~\cite{Jacksontext},
we can write this as:
\begin{equation}
\Delta E = \left(\frac{e^2}{4\pi\epsilon_0 R}\right) 
\frac{\epsilon_r -1}{2 \epsilon_r^2} F_0(a/R) \label{eq:E_ion}
\end{equation}
As the ion approaches the surface of the cluster (i.~e.~ $y \rightarrow 1$),
the first term in the expression for $F_0(y)$ dominates.  It is
easily shown that for $\epsilon_r \approx 1$ and 
as $y \rightarrow 1$, the potential becomes the
same as the `image charge' potential for an ion approaching a planar 
He--vacuum interface from
the Helium side~\cite{Poitrenaud72}.
For liquid He at 3 K, $\epsilon_r = 1.05646 $~\cite{CRC}. 
If we take $R = 5 \,$nm (which corresponds to $\sim 11,500$ He atoms),
the prefactor in the equation for $\Delta E/hc = 62$ cm$^{-1}$.  
If we compare the energy of an ion at the center of the 
cluster, $y = 0$, with one near the edge $y = 0.90$, we get
a wavenumber difference of $151$ cm$^{-1}$,
which is very large compared to $k_b\,T_c/hc = 0.26$ cm$^{-1}$
for a $^4$He cluster at $T_c = 0.38\,$ K.  
Spectroscopic studies of a number of impurities have demonstrated
that He clusters maintain themselves, by evaporation, at a 
temperature close to this value~\cite{Hartmann96,Hartmann_thesis}, 
as had previously been predicted~\cite{Brink90}.  
This potential means that
an ion will be strongly pushed towards the center of the cluster.  
Once in the center of the cluster, the thermal motion
should produce a Gaussian distribution with RMS displacement of
only 0.5 nm.  This can be compared to the size of
the `snowball' of frozen He around a positive ion, which is
known to have a radius of $\approx 0.6\,$nm~\cite{Schwarz75}.
Electrons in He form a `bubble' with a radius of 
$\approx 1.7\,$nm~\cite{Schwarz75}.  
The mechanism and thus time scale for equilibration
of an impurity center of mass motion with the internal motion of
the He cluster is presently unknown, but most likely involves exchange
of energy and angular momentum with quantized surface capillary waves, 
known as ripplons, as these
are the only He cluster modes thermally excited at this 
temperature~\cite{Brink90}.

This treatment of the effective potential for
an ionic impurity leaves out the interaction between the induced moments
created by the charge.  As long as one works in linear response theory,
and one can treat the He as a continuum, then these effects can be
included by a classical electrostatic calculation, which are presented
in an appendix.
Because the relative dielectric constant, $\epsilon_r$, is so close to one for
liquid He, this more exact treatment is in excellent agreement with the 
more approximate treatment given above.  This supports not only the
neglect of three body effects for the present case of an ionic impurity,
but also for the neutral impurities to be considered 
in a latter paper.

\subsection{Proposed Experiments}

Let us consider an ion in a He cluster of $R = 5$.  
Taylor expansion of the potential (Eq.~\ref{Eq:W}) 
around the center of the cluster
gives a harmonic force constant of
\begin{equation}
F = \left[ \frac{ e^2 (\epsilon_r - 1)}{4 \pi \epsilon R} \right]
  \left[ 1 + \frac{1}{2}\epsilon_r \right]^{-1} \frac{1}{R^2}
 = 6.5 \cdot 10^{-5}\, {\rm N}
\end{equation}
If we assume an effective mass of 45 times the mass of $^4$He,
which was found in bulk He for positive ion mobility~\cite{Poitrenaud72},
we get an effective
vibrational frequency $\nu = 2.3\,$GHz.  The
dimensionless length corresponding to this vibration is
$1.6$\, \AA.  
The vibrational transition moment for the n\,$= 0 \rightarrow 1$\, transition
is $\mu_{01} = 5.2$\,Debye. 
The zero point level corresponds to 
a root mean squared (RMS) velocity of $3.8\, {\rm m s}^{-1}$.
It will require 90 quanta of vibration to reach a 
velocity of $30 {\rm m \, s}^{-1}$, which is near the 
critical value for the onset of dissipation in bulk superfluid 
He~\cite{Rayfield64,McClintock95}.
Vibration around the center has a positive anharmonicity with
$x = 6.4 \cdot 10^{-4}$.  This implies that 
the $89 \rightarrow 90$ vibrational transition will be blue
shifted by $\approx 11\%$.
Even for this high level of excitation, the classical
turning point for the vibration is $\approx 1.5$\,nm, and
thus the ion will remain localized close to the center of
the cluster, and thus away from the surface, where
the approximations made in this paper are expected to break down.

This suggests the following experiment to measure the cluster size
dependence of the critical velocity.  For ion motion
with a peak classical value below the critical velocity, one
expects little damping of the vibrational motion.  Thus, excitation
of the vibrational transition should have little or no observable effects. 
However, once the vibrational velocity of the ion 
exceeds the critical velocity, the motion will become
strongly damped and the cluster will continue to absorb energy
from a microwave field.  This will lead to evaporation
of He atoms which can be detected by a mass selected and
mass analyzed beam.  By chirping the microwave
frequency, one can exploit the anharmonicity to 
produce an almost pure number state of the motion, at least
before damping becomes important, with the level of
excitation determined by end point of the chirp.  
It is easily verified that due to the large 
transition dipole moments, only modest microwave power
is required to drive the ion up to high levels of 
excitation.  The motion of positive ions
in bulk He is associated with vortex rings, which have the curious
property that their velocity is inversely proportional to their 
energy~\cite{Rayfield64}. The effect of confining the ion in a nanoscale
He cluster could dramatically alter the dynamics of ion motion.

Even below the critical velocity, one may expect some coupling between the
motion of the ionic impurity and the internal degrees of freedom of the
He cluster, particularly the surface ripplons, which are low frequency motions.
Using the methods described in the paper that follows this one,
on the potential of a neutral impurity in a large He cluster, it is
possible to estimate the size of the coupling of the ionic motion
to the ripplons.  These couplings will be expected to cause 
perturbations of the frequency of center of mass motion of
the ion as a function of excitation level and/or cluster size,
as the ion motion and ripplons pass through resonance conditions.  
It is easily seen that the lowest order in the interaction energy between the
ion displacement, $a$, and a displacement of a ripplon of 
angular momentum $L$, $S(L)$, is proportional to:
\begin{equation}
H_{\rm ripplon, ion} \approx \left(\frac{e^2}{4\pi\epsilon_0 R}\right) 
\frac{\epsilon_r -1}{2 \epsilon_r^2}\, (a/R)^L\, S(L)
\end{equation}
This and higher order coupling terms will allow energy to flow
from the ion motion to the ripplons.  Energy in the
ripplons, in turn, can lead to evaporation and observable
reduction in the size of a mass selected cluster.  
Such experiments could provide the first measurements
of the excitation levels of nanometer scale He clusters.

Another potential experiment is to examine the resonances of electrons
in He clusters.  Such solvated electrons are metastable, and 
the electrons will eventually be expelled from the cluster~\cite{Ancilotto95a}.  
Kim, Yurgenson and Northby~\cite{Kim97} 
have observed by spontaneous and infrared induced
electron detachments from large He clusters several milliseconds after
formation.  The stability of such charged clusters to electric fields
demonstrates that the electrons are in bubble states of
the He cluster, not attached as
surface states~\cite{Northby94a}.
Excitation of the center of mass motion of these `bubble states' should 
result in a dramatic increase in the rate of electron evaporation from the
cluster.  The potential for the negatively charged
bubble should be the same as for a positive ion, as long as the center
of the bubble is further from the surface than its radius, $\approx 1.7$\,nm.
The primary difference of the bubble compared to the positive ion is that
the effective mass is $\approx 243\,$ 
times the mass of $^4$He~\cite{Potrenand74}, which means
that for a 5 nm cluster, the harmonic vibrational frequency is $\approx 1\,$GHz.
Since the Harmonic vibrational frequency is proportional to the inverse
square root of the cluster radius, it should be possible to selectively
neutralize all clusters below a certain size by sweeping the microwave field
from a certain frequency to higher values.  By using microwave double
resonance experiments, it should be possible to determine the
homogeneous width of the resonance, and thus any drag that the 
electron bubble inside the cluster
may experience.

\section{Summary}

This paper has developed the potential that governs the motion of an
ion inside a nm scale He cluster.  This motion is determined by long 
range electrostatic interactions, and thus can be calculated without 
having to deal with the much more difficult problem of the energetics
of solvation of the ion in liquid He.  
It is found that despite the small dielectric constant of He, the
potential is very effective at driving the impurity ion into the center
of the cluster.  In thermal equilibrium, RMS displacements of only a few
\AA\ are expected. 

After ionization by electron impact, the He ion undergoes a `random walk'
due to charge exchange with other He atoms until it becomes localized
as a He$_2^+$ core which is then solvated~\cite{Halberstadt98}.  
The electrostatic potential
derived here will strongly bias this `random walk', given the low
temperatures of the clusters~\cite{Callicoatt96}.  
Such a bias could be expected to effect
the dynamics following ionization, including the probability of charge
exchange with an impurity, which will also likely be localized near the
center of the cluster.  

The trapping potential of an ion is found to be highly harmonic, with a 
frequency in the low microwave region, and having very large transition dipole
moments.  Experiments are suggested that would exploit these resonances,
and the ability to drive the ion up to high quantum states by an 
adiabatically sweep microwave pulse.

\section*{Acknowledgments}

This work was carried out with support 
from the Air Force High Density Materials program and 
the National Science Foundation.
The authors would like to acknowledge Milton Cole 
for a critical reading of a draft version of this paper
and for bringing several important references to our attention.
The hospitality of JILA, where the
work was completed and the paper written, is also acknowledged.

\section*{Appendix: Exact Electrostatic Treatment for Ion in He cluster}

In this appendix, the classical electrostatic energy that determines
the motion of an ion inside a He cluster will be derived, modeling the He
cluster as a dielectric sphere.  Given the errors in the previous 
open literature, it is important that the expressions given below be 
justified.  Further, comparison of the exact and approximate energy expression
given by Eq.~\ref{eq:E_ion} is important since it allows for an estimate to
be made of the approximations used in the companion paper for the
energy of neutral impurities in He, where a treatment including
many body effects is not practical. 

Outside of the cluster, we must satisfy Laplaces equation and we must have
only terms which go to zero at infinity.  Thus:
\begin{equation}
V_o(r,\theta) = \sum_{n=0}^{\infty} B_n r^{-(n+1)} P_n(\cos \theta)
\end{equation}
Inside the cluster, we have a field due to a point charge (screened by
the dielectric) plus fields due to surface charges.  The field of the later
must also satisfy Laplaces equation, but with positive powers:
\begin{equation}
V_i(r,\theta) = \frac{e}{4 \pi \epsilon} \frac{1}{|\vec{r} - \vec{a}|}
	+ \sum_{n=0}^{\infty} A_n r^n P_n(\cos \theta)
\end{equation}
to determine the angular dependence of the first term, we use the 
expansion:
\begin{equation}
\frac{1}{|\vec{r} - \vec{a}|} = \sum_{n=0}^{\infty} \frac{r_{<}^n}{r_{>}^{n+1}}
P_n (\cos \theta)
\end{equation}
where $r_{<} =$ min(r,a) and $r_{>} =$ max(r,a).  
We will be primarily interested in the potential for $r \geq a$
\begin{equation}
V_i(r \geq a,\theta) =  \sum_{n=0}^{\infty} \left[ \frac{e}{4 \pi \epsilon r}
\left( \frac{a}{r} \right)^n + A_n r^n \right] P_n (\cos \theta)
\end{equation}
We can determine $A_n$ and $B_n$ by satisfying the boundary conditions
$V_o(R,\theta) = V_i(R,\theta)$ and $\epsilon_o E_{o,r} = \epsilon E_{i,r}$.
For these equations to hold for all $\theta$, 
the respective coefficients of each
$P_n$ must be equal.
This leads to the following expressions:
\begin{equation}
A_n = \left( \frac{e}{4 \pi \epsilon_0} \right) \left( 
\frac{\epsilon_r - 1}{\epsilon_r} \right) R^{-(2 n + 1)} 
\left( \frac{n+1}{n+1+n \epsilon_r} \right)
a^n
\end{equation}
\begin{equation}
B_n = \left( \frac{e}{4 \pi \epsilon} \right) 
\left[ \frac{(2 n + 1) \epsilon_r}{n+1+n \epsilon_r} \right]
a^n
\end{equation}

We can now calculate the electrostatic energy by using the expression
given in Eq.~4.83 of Jackson~\cite{Jacksontext}:
\begin{equation}
W = \frac{1}{2} \int \rho(x) V(x) d^3x
\end{equation}
where $\rho(x) = e \delta (\vec{x} - \vec{a})$ is the free charge density.
Since our reference energy is that of a point charge in an
infinite dielectric, we must subtract off the monopole term in the
potential.  This gives:
\begin{equation}
W = \frac{e}{2} \sum_{n = 0}^{\infty} A_n a^n
\end{equation}
Putting in the above equation for $A_n$ we get:
\begin{equation}
W = \frac{1}{2} \left[ \frac{e^2 (\epsilon_r -1)}{4 \pi \epsilon R} \right]
 \sum_{n = 0}^{\infty} \left[ \frac{n+1}{n+1+n\epsilon_r} \right] 
\left(\frac{a}{R}\right)^{2n} \label{Eq:W}
\end{equation}
This potential is exactly one half that previously reported by
Antoniewicz {\it et al.}~\cite{Antoniewicz82,Ballester86}.
These authors reference a classic text~\cite{Smythe68} for their
expression, but unfortunately, the present authors could not locate a
copy of this book.  

In the limit that $\epsilon_r \rightarrow 1$, Eq.~\ref{Eq:W} agrees exactly
with the Taylor expansion of the expression given above (Eq.~\ref{eq:delE_ion})
for the energy obtained by `adding up' the ion--induced dipole energy
contributions.   
Since this earlier expression correctly gives the 
correct planer `image charge' potential as $y \rightarrow 1$, it cannot be
in error by a factor of two, and thus the error must lie with the expression
given by Antoniewicz, Ballester {\it et al.}  The Ph.D. thesis of 
Ballester~\cite{Ballester_thesis} gives the correct power series
expression without the factor of two error. 
The correct expression was also given in the Ph.D. thesis of 
Kim~\cite{Kim_thesis},
who corrected some other minor errors in Ballester's expressions.

For the case of He clusters, if the final sum is dropped in Eq.~\ref{Eq:W}, the
resulting approximation is in error by at most few percent over the
range $0 < y < 0.9$. In this same range of $y$, Eq.~\ref{eq:E_ion} agrees with
Eq.~\ref{Eq:W} even better, with error of less than 1\%.  Thus the fractional
error of the approximate treatment that neglects many body effects is
considerably smaller than might have been predicted {\it a priori},  
$\approx (\epsilon_r - 1)$.

The power series expansion for $W$ (Eq.~\ref{Eq:W}) is 
computationally useful for
small $y$ values (say $\leq 0.5$), but convergence slows dramatically as
$y \rightarrow 1$, since the real solution diverges in that limit.  By reference
to the form of the correct solution in the limit of $\epsilon_r \approx 1$
(Eq.~\ref{eq:delE_ion}),
it is possible to subtract out the power series of the divergent parts of 
the solution.  The resulting expression is:
\begin{eqnarray}
W & = & \frac{1}{2} \left[ \frac{e^2 (\epsilon_r -1)}{4 \pi \epsilon R} \right]
\times \nonumber \\
& & \,\,\, 
\frac{1}{(1\,+\,\epsilon_r)^2} \left[ \frac{1\,+\,\epsilon_r}{1 - y^2}
+ \frac{\epsilon_r}{y} \ln \left( \frac{1+y}{1-y} \right) \right. \nonumber\\
&  & \,\,\, \left.
+ \epsilon_r ( \epsilon_r - 1) \sum_{n = 0}^{\infty}
\frac{1}{(2n+1)(n+1+n\epsilon_r)}y^{2n} \right]
\end{eqnarray}
In this form, the fact that $W$ goes exactly into Eq.~\ref{eq:delE_ion} in the
limit $\epsilon_r \rightarrow 1$ is transparent.  Further, the last
sum is convergent, even at $y = 1$, though the convergence is slow there.
However, in the region of slow convergence, the sum makes a negligible
contribution to the total energy.  Further, if the sum is truncated
at the $n = N-1$\, term, the remainder of the sum can be approximated by
an integral:
\begin{eqnarray}
\sum_{n = N}^{\infty} \frac{1}{(2n+1)(n+1+n\epsilon_r)}\,y^{2n}
&\stackrel{N \gg 1 }{\approx} \nonumber \\
\,\,\,
\frac{1}{2(1+\epsilon_r)} \int_{N+1/2}^{\infty} \frac{y^{2n'-1}}{n'^2}dn' \\
\,\,\,
 = \frac{|\ln y|}{y(1 + \epsilon_r)}\, \Gamma(-1,(2\,N+1)\,|\ln y|)
\end{eqnarray}
where $\Gamma(a,x)$ is the incomplete Gamma function.
Ballester~\cite{Ballester_thesis} suggests using the method known as
Aitkin's $\delta^2$ process~\cite{NumericalRecipes} to speed up 
convergence of the series. 
Kim~\cite{Kim_thesis} 
derived a similar closed form expression for the interaction energy by
a different method.

\end{document}